\pdfoutput=1

\documentclass[aps, prd, twocolumn, showpacs,amsmath,amssymb,superscriptaddress, nofootinbib, showkeys, 10pt]{revtex4-1}

\usepackage[hyperindex]{hyperref}
\usepackage[utf8]{inputenc}
\usepackage{amsfonts}
\usepackage{graphicx}
\usepackage{array}
\usepackage{dcolumn}

\newcommand{\mnras}{MNRAS}

\newcommand{\mscript}[1]{{\mbox{\scriptsize #1}}}

\newcommand{\p}{\partial}

\newcommand{\g}{G_e}

\newcolumntype{M}[1]{>{\centering\arraybackslash}m{#1}}

\begin{document}
\title{Will-Nordtvedt PPN formalism applied to renormalization group extensions of general relativity} 

\author{J\'unior D. Toniato} \email{junior.toniato@ufes.br}\affiliation{Departamento de F\'{\i}sica, CCE, Universidade Federal do Esp\'{\i}rito Santo, Av. Fernando Ferrari 514, Vit\'{o}ria, ES, 29075-910 Brazil}
\author{Davi C. Rodrigues} \email{davi.rodrigues@cosmo-ufes.org}\affiliation{Departamento de F\'{\i}sica, CCE, Universidade Federal do Esp\'{\i}rito Santo, Av. Fernando Ferrari 514, Vit\'{o}ria, ES, 29075-910 Brazil}
\author{\'Alefe O.F. de Almeida}\email{alefe@cosmo-ufes.org}\affiliation{Departamento de F\'{\i}sica, CCE, Universidade Federal do Esp\'{\i}rito Santo, Av. Fernando Ferrari 514, Vit\'{o}ria, ES, 29075-910 Brazil}\affiliation{Institute for Theoretical Physics, University of Heidelberg, Philosophenweg 16, D-69120 Heidelberg, Germany}
\author{Nicolas Bertini}\email{nicolas.bertini@cosmo-ufes.org}\affiliation{Departamento de F\'{\i}sica, CCE, Universidade Federal do Esp\'{\i}rito Santo, Av. Fernando Ferrari 514, Vit\'{o}ria, ES, 29075-910 Brazil}

\date{\today}

\begin{abstract}
We apply the full Will-Nordtvedt version of the Parameterized Post-Newtonian (PPN) formalism to a class of General Relativity extensions that are based on nontrivial renormalization group (RG) effects at large scales. We focus on a class of models in which the gravitational coupling constant $G$ is correlated with the Newtonian potential. A previous PPN analysis considered a specific realization of the RG effects, and only within the  Eddington-Robertson-Schiff version of the PPN formalism, which is a less complete and robust PPN formulation. Here we find stronger, more precise bounds, and with less assumptions.  We also consider the External Potential Effect (EPE), which is an effect that is intrinsic to this  framework and depends on the system environment (it has some qualitative similarities to the screening mechanisms of modified gravity theories). We find a single particular RG realization that is not affected by the EPE. Some physical systems have been pointed out as candidates for measuring the possible RG effects in gravity at large scales, for any of them the Solar System bounds need to be considered.
\end{abstract}

\maketitle

\section{Introduction}

The use of general relativity (GR) on  the very small scales of the universe, or on  the very large ones, leads to inconsistencies or to the need of unexpected new features in the universe. These issues related with GR include quantum-gravity issues \cite{Birrell:1982ix, Buchbinder:1992rb, Susskind:1994sm, ArkaniHamed:1998nn, Ashtekar:2014kba,  Modesto:2017sdr}, the cosmological dark sector \cite{Weinberg:1988cp, Capozziello:2010zz, Famaey:2011kh, Popolo:2012bna, 0521516005}, inflation \cite{Starobinsky:1980te, Nojiri:2008ku, Weinberg:2009wa, Baumann:2009ds}, and perhaps smaller details like the small scale issues of the standard cosmological model \cite{Rodrigues:2014xka, Lelli:2015wst, Lin:2016vmm, DelPopolo:2016emo, Rodrigues:2017vto, 2017NatAs...1E.121F}. On the other hand, GR has achieved great success through several tests, specially at the Solar System scale \cite{Will:2014kxa} (see however \cite{Iorio:2014roa} for some anomalies). 

Here we analyse a class of GR extensions that is based on Renormalization Group (RG) expectations considering gravity at large distances. There are different approaches for extending  GR using RG effects, both in the high and the low energy limits \citep{Julve:1978xn, Salam:1978fd, Fradkin:1981iu, Nelson:1982kt, Goldman:1992qs, Bertolami:1993mh, Asorey:1996hz, Shapiro:1999zt,Bonanno:2000ep, Bonanno:2001hi, Reuter:2001ag, Bonanno:2001xi, Bentivegna:2003rr, Reuter:2004nx, Bonanno:2004ki,Niedermaier:2006wt, Shapiro:2009dh,Weinberg:2009wa, Bonanno:2012jy, Sola:2013fka, Modesto:2017hzl}, these also include the asymptotic safety approach to quantum gravity \citep{Niedermaier:2006wt, Percacci:2007sz, Reuter:2012id, Ashtekar:2014kba}. We consider the RGGR (Renormalization Group extended General Relativity) approach \cite{Rodrigues:2009vf, Rodrigues:2015hba}, which extends and generalizes the proposals of Refs.~\cite{Goldman:1992qs, Bertolami:1993mh, Shapiro:2004ch, Reuter:2003ca}. One of the characteristic features of this approach is the use of a correlation between the RG scale $\mu$ and the Newtonian potential in the context of stationary and weak field systems (some other proposals use $\mu \propto 1/r$, which coincides with the RGGR proposal for point particles, see also Ref.~\cite{Domazet:2010bk}). Another  RGGR feature is the use of a constant infrared $\beta$-function\footnote{A $\beta$-function of a coupling $X$ is defined by $\beta_X \equiv \mu \partial X/ \partial \mu$, where $\mu$ is the RG scale \cite{Weinberg:1996kr}. From the integration of the $\beta$-function one finds $X(\mu)$.} for the gravitational coupling constant $G$ (as explained in Ref.~\cite{Farina:2011me}, for instance). The existence of a correlation between $\mu$ and the Newtonian potential is here assumed, but this work is not restricted to a specific  form of this correlation, or to a single specific $\beta$-function.

Some different systems have been considered for evaluating large scale RG effects in gravity, from galaxies to cosmology. Considering galaxies, previous tests of RGGR in galaxies have found that the non-Newtonian effects can act as a kind of effective dark matter if $\bar \nu \gtrsim 10^{-9}$ \cite{Rodrigues:2009vf, Rodrigues:2012qm, Rodrigues:2014xka}, otherwise the effect could still be true but it would have negligible impact as a kind of dark matter, even for the smallest galaxies. The dimensionless parameter $\bar \nu$ sets the strength of the RG effects in a given system, and it is such that $\bar \nu=0$ corresponds to classical GR.

The Solar System data have always to be considered, since it provides some of the clearest and precisest results on gravity. The first work on RGGR and the Solar Syetem used the Laplace-Runge-Lenz (LRL) vector and found  $|\bar \nu_\odot| \lesssim 10^{-17}$ \cite{Farina:2011me}. A second work evaluated a number of different observations in the Solar System, to conclude  that $|\bar \nu_\odot| \lesssim 10^{-21}$ \cite{Zhao:2015pga}. A third work used up to date data on LRL vector and the Eddington-Robertson-Schiff PPN formulation to find $|\bar \nu_\odot| \lesssim 10^{-16}$ \cite{Rodrigues:2016tfm}. The third work express  the best bound on $\bar \nu$ at the Solar System up to this paper. This bound is softer than the others since the corresponding paper was the first to notice and  consider an effect that depends on the system environment and it is part of RGGR: the external potential effect (EPE). Qualitatively, this effect can be explained as follows, for a non-relativistic system: the higher is the value of the Newtonian potential due to matter outside the system, the lower are the non-Newtonian effects inside the system. This effect was not imposed as an additional feature, it was already present in the theory, but it was neglected in previous Solar System analyses. It has superficial similarities with  the screening mechanisms of modified gravity \cite{Vainshtein:1972sx, Khoury:2003aq, Hinterbichler:2010es, Koivisto:2012za}. It is also important to stress that all these bounds consider a particular form for the $\beta$-function of $G$, and a particular form for the scale setting. Although they are natural options, they constitute nonetheless additional hypothesis that will not be necessary to the main results of this paper.

Here we apply the full Will-Nordtvedt version of the Parameterized Post-Newtonian (PPN) formalism (see \cite{Will:1993ns, Will:2014kxa} for a review) to a class of RG extensions of GR that includes RGGR as a particular case. The development of the PPN formalism relied and still relies on many researches, took many years and has currently some different bifurcations (e.g., \cite{Nordtvedt:1970uv, Will:1972zz,Blanchet:1989fg,Damour:1990pi, Klioner:1999cv,Blanchet:2003gy, Turyshev:2008dr, Blanchet:2013haa, Hohmann:2015kra, Sanghai:2016tbi}). Probably the most well known and simplest version is the Eddington-Robertson-Schiff PPN formulation, which essentially depends on two parameters, $\gamma$ and $\beta$. These  are derived from theory solutions considering a massive point within a static and spherically symmetric space-time (for a review, see e.g. \cite{1972gcpa.book.....W}). This massive particle would represent the Sun and test particles would be in place of the planets or photons. The first parameter can be found from the measurement of light bending, while the second from  Mercury's orbit precession. This simpler PPN formulation was applied to RGGR in Ref.~\cite{Rodrigues:2016tfm}. Nonetheless, a theory whose values of $\gamma$ and $\beta$ are compatible with observations may be incompatible with other experiments. Also, the use of static spherically symmetric space-time and ``point'' particles are just rough approximations which in general are not irrelevant for post-Newtonian dynamics \cite{Will:1993ns}. The Will-Nordtvedt version considers more tests, it depends on 10 parameters (nine of them are observationally constrained) and it is based on fluids, not on particles.

This work is organized as follows: in the next section we present a review on RGGR that focuses on its main features that are important for the PPN evaluation. The review includes the original noncovariant formulation and the newer covariant version. Section \ref{sec:ppn apply} briefly reviews a few essential PPN features and apply them to RG extensions of GR. In Sec.~\ref{sec:ppn parameters} the observational bounds are determined for any case in which $G$ is an analytical function of the Newtonian potential. The latter section also considers the EPE and two particular classes of $\beta$-functions, one of them being the RGGR case. Our conclusions are presented in Sec.~\ref{sec:conclusions}. In the appendices \ref{app:covariant} and \ref{app:lambda} considerations on the covariant formulation are presented, and it is shown that the standard Will-Nordtvedt PPN formalism cannot be applied to the full covariant formulation.

\section{Large scale RG extensions of classical GR} \label{sec:review}

\subsection{A brief review on RGGR and a larger class of theories}
Not all RG extensions of GR at large scales have an effective action that captures all the dynamical information, without the need of imposing field equations that are outside the action. For instance, as classified in Ref.~\cite{Reuter:2003ca}, some consider the RG improved equations (e.g., \cite{Borges:2007bh, Sola:2013fka, Lima:2014hia, Sola:2017znb}), in which the coupling constants are promoted to running ones at the level of the field equations. In the case of RGGR, this promotion is done at the level of the action, and hence it is a case of RG improved action (see e.g., \cite{Reuter:2003ca, Reuter:2004nv,  Koch:2010nn, Cai:2011kd, Domazet:2012tw, Koch:2014joa}). Moreover, it has an action that leads to all the field equations \cite{Rodrigues:2015hba}. At the action level, RGGR depends explicitly on the dimensionless constant $\nu$, which is such that $\nu=0$ corresponds to classical GR (i.e., the $\beta$-function of the gravitational coupling becomes zero).

According to the RGGR action proposal of Ref.~\cite{Rodrigues:2015hba}, which extends and generalizes the proposals in Refs.~\cite{Reuter:2003ca, Shapiro:2004ch, Reuter:2004nx, Rodrigues:2009vf}, large scale RG effects can be  described by an effective action which reads (using $c=1$),
\begin{equation}
    \label{eq:RGNOexterConst}
    S =  \int \left[ \frac{R - 2 \Lambda\{\mu\}}{16 \pi G(\mu)}  + \lambda \left( \mu - f(g, \gamma,\Psi)\right)\right] \sqrt{-g} \,  d^4x + S_{m} \, .
\end{equation}
In the above,  $S = S[g,\gamma, \mu,\lambda,\Psi]$, $S_m = S_{m}[g,\Psi]$, $\Psi$ stands for any matter fields of any nature, and $\mu$ is the RG scale, whose relation to all the other fields is stated in the action in a constraint-like way, as imposed by the Lagrange multiplier $\lambda$. The field $\gamma_{\alpha \beta}$ is called the reference metric, it only appears inside $f$, without derivatives, and its variation at the action level ensures energy-momentum conservation \cite{Rodrigues:2015hba}.  The scalars $G$ and $\Lambda$ depend on the RG scale $\mu$, but in different ways. Namely, $G$ is a standard function of $\mu$, which is fixed at the action level. The relation between $\Lambda$ and $\mu$ is not fixed at the action level, but it can and must be derived from the field equations; equivalently, this means that the corresponding $\beta$-function of $\Lambda$ is not universal, and it depends on the matter fields. Examples on how to derive $\Lambda$ for different systems  can be found in Ref.~\cite{Rodrigues:2015hba}. This system-dependent relation between $\Lambda$ and $\mu$ is stressed by the use of a different notation, namely $\Lambda\{\mu\}$ instead of $\Lambda(\mu)$. In essence this implies that a local analysis of $\Lambda$ cannot determine its global behaviour, and  that $\Lambda$ is not in general an analytical function of  $\mu$.

Before proceeding, a comment on the nature of $\gamma_{\alpha \beta}$ and background independence is in order. The splitting of the spacetime metric into a background plus quantum corrections is a convenient procedure that is largely used in the context of unveiling RG effects in  gravity. Nonetheless, it is expected that the physical phenomena uncovered from this splitting does not depend on the chosen background, that is, the RG effects  should be background independent \cite{Reuter:2008qx, Becker:2014qya, Morris:2016nda, Labus:2016lkh}. Considering the $f$ function as proposed in Ref.~\cite{Rodrigues:2015hba}, see also Appendix \ref{app:covariant}, the dependence on the reference metric $\gamma_{\alpha \beta}$ is such that, when the metrics coincide, the RG effects become null and one recovers classical GR.\footnote{More precisely,  if, in a given neighborhood of a spacetime point  the metrics satisfy $g_{\alpha \beta} = \gamma_{\alpha \beta}$, then in that neighborhood there are no RG effects.} Thus, $\gamma_{\alpha \beta}$ sets a background for the RG effects. Formally, the action (\ref{eq:RGNOexterConst}) is background independent in the sense that there is no particular geometry that is preferred. On the other hand, different coices of  $\gamma_{\alpha \beta}$ at the level of the field equations lead to different solutions for the spacetime metric. Therefore, in the sense of the split symmetry, as discussed for instance in Ref.~\cite{Reuter:2008qx}, action \eqref{eq:RGNOexterConst} is not explicitly background independent. This does not imply that $\gamma_{\alpha \beta}$ is a physically independent quantity, only that action \eqref{eq:RGNOexterConst} does not handle background changes (i.e., changes of $\gamma_{\alpha \beta}$ within a fixed coordinate system). In classical GR, sometimes the boundary conditions are not obvious and one has to use physical intuition (or large computational efforts) to discover the proper boundary conditions. In the case of the RG extended action \eqref{eq:RGNOexterConst}, this problem includes finding the proper reference metric.\footnote{Actually the problem is much simpler, since for the proposed $f$ function one only has to specify a scalar quantity, $u^\alpha  u^\beta \gamma_{\alpha \beta}$ (see Appendix \ref{app:covariant}).} For systems that are close to Newtonian, the most natural assumption for $\gamma_{\alpha \beta}$ is the Minkowski metric, in this case the RG effects depend on the Newtonian potential, and this is the choice assumed in this work. This choice is relevant for the passage from the covariant action \eqref{eq:RGNOexterConst} to the noncovariant RGGR formulation (further details are in Appendix \ref{app:covariant} and in Refs.~\cite{Rodrigues:2015hba, Rodrigues:2016tfm}).

Considering the field equations, from  action \eqref{eq:RGNOexterConst}, the variation with respect to $\lambda$ yields the scale setting $\mu = f$, the variation with respect to $\mu$ yields a condition between $G, \Lambda, \lambda$ which ensures that the matter energy-momentum tensor satisfies\footnote{Due to the constraint term that depends on both $\mu$ and the matter fields, the diffeomorphism invariance of $S_m$ is not sufficient to assure that $T_{\mu\nu}$ is conserved \cite{Rodrigues:2015hba,Wald:1984rg}.}  $\nabla_\alpha T^\alpha_\beta \propto \lambda$. The variation with respect to $\gamma_{\alpha \beta}$ sets $\lambda = 0$ at the level of the field equations (whenever $\partial f / \partial \gamma^{\alpha \beta} \not=0$), thus ensuring energy-momentum conservation (see Ref.~\cite{Rodrigues:2015hba}  for further details). At last, the variation with  respect to the metric yields 
\begin{equation}
    \label{eq:fieldext}
    {\cal G}_{\alpha \beta} + \Lambda g_{\alpha \beta} = {8 \pi G} T_{\alpha \beta},
\end{equation}
where 
\begin{equation}
        {\cal G}_{\alpha \beta} \equiv G_{\alpha \beta} +  g_{\alpha \beta} G \Box G^{-1}  - G \nabla_\alpha \nabla_\beta G^{-1},
\end{equation}
$\Box \equiv g^{\alpha \beta} \nabla_\alpha \nabla_\beta$, and $\nabla_\alpha$ is the  covariant derivative.

If it is possible to neglect the contribution from $\Lambda$  in the Solar System up to first post-Newtonian order, which seems natural since $\Lambda$ should be a correction to the cosmological constant $\Lambda_0$ that appears in classical GR, it is useful to write eq.~\eqref{eq:fieldext} as
\begin{align}\label{eqd}
R_{\alpha\beta}= G\bigg[8\pi & \left(T_{\alpha\beta}-\frac{1}{2}g_{\alpha\beta}\,T\right) \ + \nonumber \\
& + \ \nabla_\alpha \nabla_\beta \left(G^{-1}\right) + \frac{1}{2}g_{\alpha \beta}\Box\left(G^{-1}\right)\bigg]\,.
\end{align}

The $\Lambda$ term was not considered in the previous Solar System analysis of RGGR \cite{Farina:2011me, Zhao:2015pga, Rodrigues:2016tfm}. In Appendix \ref{app:lambda} we comment on the possible effects of $\Lambda$ and show that the Will-Nordtvedt PPN formalism in its standard form cannot handle the $\Lambda$ term, in particular because $\Lambda$ cannot be both an analytical function and be compatible with asymptotic flatness.

\subsection{Running gravitational constant and scale setting}
For concreteness, it helpful to present an example for the $G(\mu)$ function. We present below a simple expression for $G(\mu)$ that some of us used in previous publications, and which was also derived from different RG approaches, namely \cite[e.g.,][]{Fradkin:1981iu, Nelson:1982kt, Reuter:2003ca,  Shapiro:2004ch, Bauer:2005rpa, Farina:2011me},
\begin{equation}
    \label{eq:Gmu}
    G(\mu) = \frac{G_0} {1 + 2 \nu \ln (\mu/\mu_0)}\, ,
\end{equation}
where $G_0$ and $\mu_0$ are  constants such that $G(\mu_0) = G_0$, and $\nu$ is a small dimensionless constant. GR is recovered with  $\nu=0$. We present further details on the consequence of this expression in Sec.~\ref{sec:ppn parameters}, but our main results in this work are not limited to this expression. 

From the action (\ref{eq:RGNOexterConst}), the relation of the scale $\mu$ to other physical quantities (i.e., the scale setting \cite{Babic:2004ev, Domazet:2010bk}) is a field equation that comes from the variation of the action with respect to the Lagrange multiplier $\lambda$. Contrary to some other approaches, the scale setting is not an additional equation outside the action, it is derived from the action. In Ref.~\cite{Rodrigues:2009vf}, considering RG expectations within GR at large scales, some of us proposed that, within stationary weak field gravitational systems, there should be a function $f$ such that, in a given reference frame,
\begin{equation} \label{eq:ss}
	\mu = f(U) \, .
\end{equation}
Our results do not depend on specifying a particular $f$ function. In the above, $U$ is the negative of the Newtonian potential\footnote{For conciseness, commonly we will call $U$ the Newtonian potential, without writing ``negative'' in front of it. We use $U$ since we are following the notation of Ref.~\cite{Will:1993ns} on the PPN parameters and the potentials.} \cite{Will:1993ns}, and it is given by,
\begin{equation}\label{u}
	U(t, \vec x) = G_\mscript{N} \int \frac{\rho(t, \vec x \, ')}{| \vec x - \vec x \, '| } d^3x' \, ,
\end{equation}
where $G_\mscript{N}$ is the Newtonian gravitational constant (which may be different from $G_0$). The scale setting \eqref{eq:ss} is a development over some previous RG application to gravity at large scales, in which it was used the qualitative relation $\mu \sim 1/r$ \cite[e.g.,][]{Shapiro:2004ch, Reuter:2004nx}, that is, a large value for the RG scale $\mu$ should correspond to small distances $r$. The use of the Newtonian potential is in qualitative agreement with the previous assumption, but it includes a dependence on the mass distribution and uses the most relevant potential for systems close to the Newtonian regime, that is, $U$.

Since $U$ is not a spacetime scalar, the scale setting (\ref{eq:ss}) is not covariant. In Refs.~\cite{Rodrigues:2015hba, Rodrigues:2016tfm} we proposed a covariant generalization of the above scale setting, but we leave further details on the covariant version to the Appendix \ref{app:covariant}.  As shown in the latter appendix, the covariant version leads to the appearance of potentials that are not part of the Will-Nordtvedt PPN formalism.

In Ref.~\cite{Rodrigues:2016tfm} we used eq.~(\ref{eq:Gmu}) and presented a natural $f(U)$ function which lead to a metric solution with spherical symmetry that could be handled through the Eddington-Robertson-Schiff PPN formalism. Here we will proceed with more generality,  namely we will simply demand that $G$ can be expanded as a function of $U$ as follows,
\begin{equation}\label{eq:GU}
G^{-1}(\mu) =  G^{-1}(U)= G^{-1}_{e} + 2\sum_{n=1}^\infty \nu_n U^n.
\end{equation}
With this parametrization, GR is recovered with $\nu_n =0$. It will be shown that all the $\nu_{n}$ terms with $n \geq 3$ are not relevant to the Solar System dynamics up to the first post-Newtonian order. In eq.~\eqref{eq:GU}, $\nu_n$ are real constants and $G_e$ is the value of $G(U)$ when $U=0$. We use the index $e$ in reference to the external value of $G$. That is, far from the Sun, the Newtonian potential of the Solar System should become close to zero, but the value of $G$ at such distance may depend on the environment of the Solar System. This will be further developed in Sec.~\ref{sec:ppn parameters} and the relation between $G_e$ and $G_\mscript{N}$ is shown in the next section.

\section{The post-Newtonian approximation} \label{sec:ppn apply}
In this section we apply the Will-Nordtvedt PPN formalism  \cite{Will:1993ns, Will:2014kxa} to a RG extension of GR whose RG scale $\mu$ is correlated with the Newtonian potential. This formalism uses a perfect fluid as the gravitational source and describes the metric of a gravitational theory in terms of ten observable PPN parameters in a theory-independent way. The main small parameter of the formalism is the velocity field $|\vec v| = v < 1$. The metric is expanded about Minkowski spacetime,
\begin{equation} \label{eq:getah}
g_{\alpha\beta}=\eta_{\alpha\beta}+h_{\alpha\beta}\,,
\end{equation}
where  $\eta_{\alpha\beta}$ is the Minkowski metric, which is of zeroth-order on $v$, and $h_{\alpha \beta} \sim O(v^2)$, at least. We use the signature $(-,+,+,+)$.

Up to the first post-Newtonian order, the metric must be known as follows: $g_{00}$ to order $v^4$, $g_{0i}$ to order $v^3$ and $g_{ij}$ to order $v^2$ (Latin indices run from $1$ to $3$). Thus, up to the required order, the Ricci tensor components can be expressed as 
\begin{align}
R_{00} =& -\frac{1}{2}\nabla^2h_{00} - \frac{1}{2}\left(h^k_{~k,00}- 2\,h^k_{~0,k0} \right) - \frac{1}{4}\,|\vec{\nabla}h_{00}|^2 \ +\nonumber\\[1ex]
&+\frac{1}{2}\,h_{00,l}\left(h^{lk}_{~~,k}- \frac{1}{2}\,h^k_{~k,j}\delta^j_l\right)+ \frac{1}{2}\,h^{kl}h_{00,lk}\,,\label{1}\\[2ex]
R_{0i}=& -\frac{1}{2}\left(\nabla^2h_{0i} - h^k_{~0,ik} + h^k_{~k,0i} - h^k_{~i,k0} \right)\,,\label{2}\\[2ex]
R_{ij}=& -\frac{1}{2}\!\left(\nabla^2h_{ij} -\! h_{00,ij} +\! h^k_{~k,ij}-\! h^k_{~i,kj}- \! h^k_{~j,ki} \right).\label{3}
\end{align}
The comas refer to simple derivatives, $\nabla^2 \equiv \eta^{ij}\p_i\p_j$\,, and it was used that time derivatives effectively count as one order increase. Thus, if a quantity $X$ is of order $v^n$ then $X,_k\sim O(v^n)$ and $X,_0\sim O(v^{n+1})$.

Using eq.~\eqref{eq:GU} and that  $U \sim O(v^2)$ for systems not far from equilibrium, then
\begin{eqnarray}
\nabla_\alpha \nabla_\beta \left(G^{-1}\right)  & =  & (G^{-1})_{, \alpha \beta} - \Gamma_{\alpha \beta}^\lambda (G^{-1})_{, \lambda} \, , \nonumber \\[.1cm]
&=&   2\nu_1  \left( U,_{\alpha \beta}  -  \Gamma^{\lambda}_{\alpha\beta}U,_{\lambda}  \right) +  \\
&&  + \;  4\nu_2 \big(U,_{\alpha}U,_{\beta} + UU,_{\alpha \beta }\big) + O(v^6)\,. \nonumber
\end{eqnarray}

Since the gravitational source is  a perfect fluid, 
\begin{equation}\label{emt}
T^{\mu\nu}=(\rho+\rho\Pi+p)u^\mu u^\nu + pg^{\mu\nu}\,,
\end{equation}
where $\Pi$ is the specific energy density, $p$ is the pressure and $u^\mu=(u^0,v^i)$ is the four velocity of the fluid element, with
\begin{equation}\label{u0}
u^0= \sqrt{\frac{1+v^2}{1-h_{00}}}\,,
\end{equation}
such that $u_\mu u^\mu=-1$. The mass density $\rho$, $\Pi$ and $p/\rho$ are of order $v^2$ \cite{Will:1993ns}.

With the expressions above, we compute the metric components order by order on powers of $v$.

\paragraph{$h_{00}$ up to order $v^2$ (Newtonian limit):}
Up to the required order,
\begin{equation}
R_{00}=-\frac{1}{2}\,\nabla^2h_{00} \quad \mbox{and} \quad T_{00}=-T=\rho\,.
\end{equation}
Therefore,
\begin{equation}
\nabla^2h_{00}=-8\pi \g \rho + 2\g\nu_1 \nabla^2U
\end{equation}
and, from eq.~\eqref{u},
\begin{equation} \label{eq:h00GU}
h_{00}=2\frac{\g}{G_\mscript{N} }(1+G_\mscript{N} \nu_1)U\,.
\end{equation}

In order to be in agreement with the Newtonian physics,
\begin{equation}\label{newtlimit}
h_{00}=2U\,,
\end{equation}
thus  we must set
\begin{equation} \label{eq:GeGN}
\g(1+G_\mscript{N} \nu_1)=G_\mscript{N} \,.
\end{equation}
The equation above sets the relation between $\g$ and $G_\mscript{N}$. Since this relation is now clear, henceforth we  use 
\begin{equation}
	G_\mscript{N} =1 \, .	
\end{equation}
Thus,
\begin{equation}\label{gt}
\g=\frac{1}{1+\nu_1}\,.
\end{equation}

\paragraph{$h_{ij}$ up to order $v^2$:} 
Imposing the three gauge conditions,
\begin{equation}\label{gaugehij}
h^\mu_{~i,\mu}-\frac{1}{2}h^\mu_{~\mu,i}=2\g\nu_{1} U_{,i}\,,
\end{equation}
the spatial part of eq.~\eqref{eqd} reduces to,
\begin{equation} 
\nabla^2h_{ij}= -8\pi\g\rho\delta_{ij} - 2 \g\nu_{1} \nabla^2U\delta_{ij}\,.
\end{equation}
The above equation is easily integrated,
\begin{equation}\label{hij}
h_{ij}=2\left(1-\frac{2\,\nu_{1}}{1+\nu_{1}}\right)U\,\delta_{ij}\,,
\end{equation}
where eq.~\eqref{gt} was used.

\paragraph{$h_{0i}$ up to order $v^3$:}
With a fourth gauge condition,
\begin{equation}\label{gaugeh0i}
h^\mu_{~0,\mu}-\frac{1}{2}h^\mu_{~\mu,0}=-\frac{1}{2}h_{00,0} +3\g\nu_{1} U_{,0}\, 
\end{equation}
and from eq.~\eqref{eqd},
\begin{equation} \label{eqh0i}
\nabla^2h_{0i} + \g U_{,0i}= 16\pi\g\rho v_i\,.
\end{equation}
To integrate the above equation, we will use the super-potential $\chi(t,\vec{x})$ \cite{Will:1993ns}, which is given by
\begin{equation}\label{superpotential}
\chi(t,\vec{x})\equiv \int{\rho(t,\vec{x}')|\vec{x}-\vec{x}'|\,d^3x'}\,.
\end{equation}
From the above definition,
\begin{equation}
\nabla^2\chi=-2U\, \quad \mbox{and} \quad \chi_{,0i}=V_i-W_i\,,
\end{equation}
where
\begin{equation}
V_i=\int{\frac{\rho(t,\vec{x}')\,v_i'}{|\vec{x}-\vec{x}'|}\,d^3x'}\, , \quad \nabla^2V_i=-4\pi\rho v_i\,,
\end{equation}
and
\begin{equation}
W_i=\int{\frac{\rho(t,\vec{x}')\vec{v}\cdot (\vec{x}-\vec{x}')(x-x')_i}{|\vec{x}-\vec{x}'|^3}\,d^3x'}\,.
\end{equation}
Therefore, from eq.~\eqref{eqh0i} it results
\begin{equation}\label{h0i}
h_{0i}= -\frac{7V_i}{2(1+\nu_{1})} - \frac{W_i}{2(1+\nu_{1})}\,.
\end{equation}

\paragraph{$h_{00}$ up to order $v^4$:}
To develop the right hand side of the dynamical equation, we need the explicit expression of some components of the connection. To the required order, that terms are
\begin{align}
\Gamma^i_{00}=& -U_{,i}\,, \\[1ex]
\Gamma^k_{ij}=& \left(1-\frac{2\,\nu_{1}}{1+\nu_{1}} \right)\left(U_{,i}\delta^k_{j}+ U_{,j}\delta^k_{i}- U^{,k}\delta_{ij} \right).
\end{align}
For the energy-momentum tensor, up to order $v^4$, one finds
\begin{equation}
T_{00}-\frac{1}{2}\,g_{00}T=\frac{1}{2}\,\rho\left[1+2\left(v^2-U+\frac{\Pi}{2} + \frac{3p}{2\rho} \right) \right]\,,
\end{equation}
where the expansion of eq. \eqref{u0} was used. By considering the gauge fixing conditions \eqref{gaugehij}, \eqref{gaugeh0i}, introducing the potentials below \cite{Will:1993ns},
\begin{align}
&\nabla^2\Phi_1=-4\pi\rho v^2, \qquad \nabla^2\Phi_2=-4\pi\rho U, \nonumber\\[1ex]
&\nabla^2\Phi_3= -4\pi\rho \Pi\,, \qquad \nabla^2\Phi_4=-4\pi p,
\end{align}
and using the relation,
\begin{equation}
|\vec{\nabla}U|^2=\nabla^2\left(\frac{U^2}{2}-\Phi_2\right) \, ,
\end{equation}
the dynamical equation can be integrated, leading to
\begin{align} 
h_{00}=& \ 2U -2\left[1+ \frac{\nu_{1}^2-\nu_{2}(1+\nu_{1})}{(1+\nu_{1})^2}\right]\!U^2 \ +\nonumber\\[1ex]
& \ + \frac{4\Phi_1}{1+\nu_{1}} +  \frac{2\Phi_3}{1+\nu_{1}}+ \frac{6\Phi_4}{1+\nu_{1}} \ +\nonumber\\[1ex]
& \ + \left[\frac{4\left(1-\nu_{1}+\nu^2_{1}\right)}{(1+\nu_{1})^2}-\frac{4\nu_{1}}{1+\nu_{1}}\right]\Phi_2+O(v^6)\,. \label{eq:h00rggr}
\end{align}

With the above, we conclude the expansion of the RGGR perturbations as a function of the PPN potentials. In the next section, we infer the values of the PPN parameters and compare with the observational values.

\section{The PPN parameters and their interpretation} \label{sec:ppn parameters}

\subsection{General analysis} \label{sec:ppn general}
With the results obtained in the previous section, the  metric up to the first post-Newtonian (1PN) order can be written as
\begin{align}
g_{00}=& -1 +2U -2\left[1+ \frac{\nu^2_{1}-\nu_{2}(1+\nu_{1})}{(1+\nu_{1})^2}\right]\!U^2 \ +\nonumber\\[2ex]
& \  +\left(1-\frac{\nu_{1}}{1+\nu_{1}}\right)\Big(4\Phi_1 + 2\Phi_3 + 6\Phi_4\Big) \ + \nonumber \\[2ex]
& \ + 4\left[1-\frac{4\nu_{1}+\nu^2_{1}}{(1+\nu_{1})^2}\right]\Phi_2\,, \label{33} \\[2ex]
g_{0i}=& \left(1-\frac{\nu_{1}}{1+\nu_{1}}\right)\Big( -\frac{7V_i}{2} - \frac{W_i}{2}\Big)\,,\nonumber \\[2ex]
g_{ij}=& \ \delta_{ij}+ 2\left(1-\frac{2\nu_{1}}{1+\nu_{1}}\right)U\,\delta_{ij}\,.\nonumber
\end{align}
To extract the PPN parameters from the above geometric structure we compare it to the Will-Nordtvedt generic post-Newtonian metric \cite{Will:1993ns}, namely
\begin{align}
g_{00} =& -1 + 2U - 2\beta U^2 + (2 \gamma +2+\alpha_3 +\zeta _1-2 \xi ) \Phi_1 \ + \nonumber \\[1ex]
& + 2(3 \gamma -2\beta+1+\zeta _2+ \xi ) \Phi_2 +2(1+\zeta _3 ) \Phi_3 \ + \nonumber \\[1ex]
& + 2(3 \gamma +3\zeta _4-2 \xi ) \Phi_4 - (\zeta _1-2 \xi ) {\cal A}  -2\xi \Phi_W, \nonumber \\[2ex]
g_{0i} =& - \frac{1}{2}(4 \gamma +3+\alpha_1-\alpha_2+ \zeta_1-2\xi) V_i \ - \label{34} \\[1ex]
&- \frac 1 2(1+\alpha_2- \zeta_1+2\xi) W_i\,, \nonumber\\[2ex]
g_{ij} =& \ (1+2\gamma \,U)\, \delta_{ij}\,.\nonumber
\end{align}

From the coefficients of $U$ in $g_{ij}$ and $U^2$ in $g_{00}$, one infers the parameters $\gamma$ and $\beta$ as functions of $\nu_1$ and $\nu_2$. Using the data from Table \ref{tab}, one finds
\begin{eqnarray}
	|\nu_1| < 1.2 \times 10^{-5} \, , \quad	|\nu_2| < 8 \times 10^{-5}\, .\label{nu1nu2GammaBeta}
\end{eqnarray}
We stress that the above considers {\it only} the observational constraints from $\gamma$ and $\beta$, which are not all the observational constraints.

\begin{table}[ht]
	\centering 
	\caption{\label{tab} Limits on the PPN parameters, considering only the strongest limits for each parameter \cite{Will:2014kxa}. The $\zeta_4$ does not have a direct measurement. These limits apply to the absolute value of each parameter.}
		\begin{tabular}{cd}
			\toprule
			Parameter &  \multicolumn{1}{@{\hspace{3em}}c}{Limit} \\
			\hline
			$\gamma-1$& 2.3 \times 10^{-5}\\
			$\beta-1$ & 8. \times 10^{-5} \\
			$\xi$     & 4. \times 10^{-9} \\
			$\alpha_1$& 4. \times 10^{-5}\\ 
			$\alpha_2$& 2. \times 10^{-9}\\
			$\alpha_3$& 4. \times 10^{-20}\\
			$\zeta_1$ & 2. \times 10^{-2}\\
			$\zeta_2$ & 4. \times 10^{-5}\\
			$\zeta_3$ &  1. \times 10^{-8}\\
			$\zeta_4$ & \multicolumn{1}{@{\hspace{3em}}c}{---}\\
			\botrule
		\end{tabular}
\end{table}

Since $\nu_1$ and $\nu_2$ need to be much smaller than one, their relations to the PPN parameters can be expressed from linear expansions on $\nu_1$ and $\nu_2$, which reads
\begin{align}
&\gamma=1- 2\nu_1 \,, \nonumber \\
&\beta=1-\nu_2 \,, \nonumber\\ 
& \alpha_{2}=-\nu_1 \, , \nonumber \\
&\zeta_{2}=-2(\nu_1+\nu_2)\,,\label{param}\\
& \zeta_{3}=-\zeta_{4}=-\nu_1\, , \nonumber \\
&\alpha_{1}=\alpha_{3}=\xi=\zeta_{1}=0\,.\nonumber
\end{align}

Using the relations above and the observational constraints of {\it all} the the PPN parameters, listed in Table \ref{tab}, the resulting strongest constraints on the parameters $\nu_1$ and $\nu_2$ are displayed in Table \ref{tab:nu1nu2}. One sees that they do not come from $\beta$ or $\gamma$, but from $\alpha_2$ and $\zeta_2$.
\begin{table}[ht]
	\centering 
	\caption{\label{tab:nu1nu2} Strongest constraints on $\nu_1$ and $\nu_2$ from all the observational constraints on the PPN parameters.}
		\begin{tabular}{ccc}
			\toprule
			Constraint & \hspace*{1cm}& \mbox{Origin}\\
			\hline\\[-0.2cm]
			$|\nu_1| < 2\times 10^{-9}$& &\mbox{$\alpha_2$ constraint}\\[.05in]
			$|\nu_2| < 2\times 10^{-5}$& &\mbox{$\zeta_2$ constraint}\\[.05in]
			\botrule
		\end{tabular}
\end{table}

There are well known examples of theories that come from an action and have $\alpha_1$ and $\alpha_2$ different from zero, which are related with special frame effects \cite{Will:2014kxa, Clifton:2011jh}, but theories with an action are not expected to yield non-zero values for any of the $\zeta$'s and $\alpha_3$ if $\xi = 0$ \cite{Lee:1974nq}. On the other hand, we are not using the full covariant action, which demands energy-momentum conservation, but the noncovariant approximation. The derived bound from $\zeta_2$ changes the bound found from the $\beta$ parameter by a factor 4 (from eq.~\ref{nu1nu2GammaBeta}). That is, the noncovariant approximation works as an order of magnitude approximation, at the 1PN order, to the covariant version \cite{Rodrigues:2016tfm}. The situation would be  different in case  $\alpha_3$ would depend on $\nu_1$ or $\nu_2$. Further considerations on the effects from $\Lambda$ and the full covariant action are in appendices \ref{app:covariant} and \ref{app:lambda}.

\subsection{A constant infrared $\beta$-function and the External Potential Effect} \label{subsec:EPE}

There is a particular expression for $G(U)$ that is well motivated and particularly simple. This expression was proposed in Ref.~\cite{Rodrigues:2009vf} and it reads
\begin{equation} \label{eq:Gnubar}
 G^{-1} = G^{-1}_0 \left[ 1 + 2  \bar \nu \ln \left( \frac{U}{U_0} \right) \right ],	
\end{equation}
where $\bar \nu$ is a constant and $U_0$ is a reference potential (the one that satisfies $G(U_0) = G_0$). The above expression uses the following infrared $\beta$-function of $G$ \cite[e.g.,][]{Fradkin:1981iu, Nelson:1982kt, Reuter:2003ca, Bauer:2005rpa, Farina:2011me} (with $c = \hbar = 1$),
\begin{equation}
	\beta_{G^{-1}} \equiv \mu \frac{\partial G^{-1}(\mu)}{\partial \mu} = 2 \nu M^2_{\mbox{\tiny Planck}} = 2 \nu G_0^{-1},
\end{equation}
whose integration leads to $G^{-1}(\mu) = G_0^{-1}(1 + 2 \nu \ln \mu/\mu_0)$. The latter expression is combined with the scale setting \cite{Rodrigues:2009vf, Domazet:2010bk, Rodrigues:2016tfm}
\begin{equation} \label{eq:ssstandard}
	\mu = \left( \frac{U}{U_0} 	 \right)^\alpha.
\end{equation}
 In eq. (\ref{eq:Gnubar}) we used $\bar \nu \equiv \nu \alpha$.

The $G(U)$ expression from eq.~(\ref{eq:Gnubar}) is not  in general compatible with the expansion (\ref{eq:GU}), but it becomes compatible once the external potential effect (EPE) is considered \cite{Rodrigues:2016tfm}.

Since in this picture $G$ depends on the potential $U$, $G$ will in general depend on both the matter distribution inside the system under investigation and also the matter outside it. Following Ref.~\cite{Rodrigues:2016tfm}, we write,
\begin{eqnarray}
	\rho = \rho_s + \rho_e \, , \nonumber \\
	U = U_s + U_e \, ,
\end{eqnarray}
where $\rho_s$ refers to the matter density contribution that is inside the system under consideration, while $\rho_e$ refers to the external mass density. The quantities $U_s$ and $U_e$ are computed from eq.~(\ref{u}), but with  $\rho$ replaced by $\rho_s$ and $\rho_e$ respectively.

We consider that the scale of the system is much smaller than the typical scale of the exterior contributions (e.g., the Solar System inside the Galaxy), such that inside the system $U_e$ behaves as a constant. Hence, instead of using the arbitrary $U_0$ scale, it is convenient to use $U_e$ as the reference potential, as follows,\footnote{The change on the scale from $U_0$ to $U_e$ actually changes  $G_0$ to $G_e$ and also changes $\bar \nu$, such that the product $G^{-1}\nu$ is constant. The relevant change is on the reference potential, the changes on $G$ and $\nu$ are second order on $\nu$. For clarity, we opted not to introduce an index on $\nu$ to label this small change. The exact expressions can be found in Ref.~\cite{Rodrigues:2016tfm}.}
\begin{eqnarray}
	G^{-1} &=& G^{-1}_e \left[ 1 + 2  \bar \nu \ln \left( 1 + \frac{U_s}{U_e} \right) \right ], \nonumber \\[.1cm]
	&=& G^{-1}_e \left(  1 + 2 \bar \nu \frac{U_s}{U_e} -  \bar \nu \frac{U_s^2}{U_e^2} \right) +... \label{eq:GUs}
\end{eqnarray}
with $G(U_e) = G_e$ (or, equivalently, $G|_{U_s = 0} = G_e$) and $U_s < U_e$. The expression above is compatible with eq.~(\ref{eq:GU}), with $U_s$ in place of $U$. Hence, we identify,
\begin{eqnarray}
	\nu_1 &=&  \frac{\bar \nu }{G_e U_e} \, , \nonumber \\
	\nu_2 &=&  - \frac{\bar \nu }{2 G_e U_e^2}. 
\end{eqnarray}
It should be remembered that the expressions above assume $U_e>U_s$, hence the limit $U_e \rightarrow 0$ is meaningless.

The PPN bound on $\bar \nu$ depends on the value of $U_e$, and it is such that the larger is $U_e$, the softer is the bound on $\bar \nu$. Since $U_e$ is a gravitational potential, $U_e < 1$, and hence the most conservative bound on $\bar \nu$ comes from using $U_e \sim 1$, which reads
\begin{equation}
	|\bar \nu| < 10^{-9}    \mbox{ for $U_e \sim 1$}.
\end{equation}

The minimum structure outside the Solar System that it should be considered is the Milky Way, whose Newtonian potential at the Solar System position can be estimated to be about $U_e \sim 10^{-6}$ \cite{Rodrigues:2016tfm}, thus,
\begin{equation} \label{eq:bound1}
	|\bar \nu| \lesssim 10^{-17}    \mbox{ for $U_e \sim 10^{-6}$}.
\end{equation}
Beyond the Milky Way, one should consider the Local Group contribution to $U$. Since the Milky Way is already one of the two most massive galaxies of the Local Group, the other being Andromeda, the bound will not change appreciably. Beyond the Local Group there is the Virgo super-cluster, but the Local Group is not gravitationally bound to it, thus one starts to enter a domain in which cosmology becomes important, and hence  beyond the validity of the scale setting (\ref{eq:ss}). Therefore, unless there is some nontrivial cosmological contribution, eq.~\eqref{eq:bound1} is the most reasonable bound on $\bar \nu$ that can be inferred at the Solar System.

The bound that appears in eq.~\eqref{eq:bound1} is slightly stronger than the bound from Ref.~\cite{Rodrigues:2016tfm}, where it was found $|\bar \nu| \lesssim 10^{-16}$ for the same value of the external potential. The reason for the disagreement comes from that here we use all the Will-Nordtvedt parameters, and the strongest bound on $\nu_2$ is not the one from $\beta$, but from $\zeta_2$. These two bounds only differ by a factor 4, but since $8 \times 10^{-5} \sim 10^{-4}$ and $2 \times 10^{-5} \sim 10^{-5}$, the final answer has an order of magnitude of difference.

\subsection{Infrared $\beta$-function proportional to $\mu^n$ and the External Potential Effect} \label{subsec:EPE2}

Although the case of a constant infrared $\beta$-function is a natural one, here we consider another simple possibility that also appears frequently in diverse contexts,
\begin{equation} \label{eq:GnuN}
	\beta_{G^{-1}} \equiv \mu \frac{\partial G^{-1}(\mu)}{\partial \mu} = \nu \mu^n,
\end{equation}
where $n$ is a dimensionless real constant different from zero and $\nu$ is a constant. Again $\nu$ is used to set the strength of the RG effects, but for the $\beta$-function above,  $\nu$ is a dimensionful quantity.

After integrating eq.~(\ref{eq:GnuN}) and using the scale setting\footnote{One could consider $\mu=f(U)$, but for clarity we consider this simpler case.} $\mu = U$,  one finds
\begin{equation} \label{eq:Gxi}
	G^{-1}(U) = G_0^{-1} + \frac{\nu}{n }  U^n\, . 
\end{equation}
In the above, $G_0$ is an integration constant. Upon considering the presence of matter outside the system, $U$ is divided into $U_s$ and $U_e$ (the latter being a constant) and the $G$ expression can be stated as a function of $U_s$ as follows,
\begin{eqnarray}
	G^{-1} &=& G_0^{-1} + \frac{\nu}{n }  (U_s + U_e)^n\,  \nonumber \\
 &=& G_e^{-1} - \frac{\nu}{n } U_e^n + \frac{\nu}{n }  (U_s + U_e)^n\,  \\
	&= & G_e^{-1} + \nu  U_e^{n-1} U_s + \nu \frac{n-1}2 U_e^{n-2} U_s^2 + O\left( \frac{U_s^3}{U_e^3} \right)\, , \nonumber
\end{eqnarray}
where $G_e$ is defined from $G(U_s=0) = G_e$.

From the expansion above and eq.~\eqref{eq:GU}, one identifies
\begin{eqnarray}
	\nu_1 &=& \frac 12 \nu U_e^{n-1} \, ,\\
	\nu_2 &=& \nu \frac{n-1}4  U_e^{n-2} \, .	
\end{eqnarray}
As in the previous subsection, for $U_e \sim 1$, the bound comes from $\alpha_2$ and reads (using $c=\hbar = G_\mscript{N} =1$),
\begin{equation} \label{eq:boundxi1}
	|\nu| < 10^{-9}.
\end{equation}
If $U_e \ll 1$ and $n=1$, then the bound above is also valid. 

For the case $U_e \sim 10^{-6}$ (which corresponds to the contribution from the Milky Way at the Solar System), and if $n$ is not close to one, the bound becomes,
\begin{equation}
	|(n-1)\nu | \lesssim   10^{-16 + 6 n} \,.
\end{equation}
The above inequality shows that the larger is the external potential $U_e$, the softer is the bound on $\nu$, as expected. 

This example with $G$ given by eq.~\eqref{eq:Gxi} shows that, for some cases, the EPE does not  improve concordance with GR. Namely, for $n=1$ the bound on $\nu$ is given by eq.~\eqref{eq:boundxi1}, which is independent of $U_e$.

\section{Conclusions} \label{sec:conclusions}

In this work we used, for the first time, the Will-Nordtvedt PPN formalism to address Solar System bounds on a class of RG-based proposals that extend GR. This class is such that the RG scale is a function of the Newtonian potential, hence in particular it includes the RGGR proposal \cite{Shapiro:2004ch, Rodrigues:2009vf, Rodrigues:2015hba}. We also consider the External Potential Effect (EPE), which is an intrinsic effect of these proposals and which depends on the environment of the system \cite{Rodrigues:2016tfm}.

In Ref.~\cite{Rodrigues:2016tfm}, using a more heuristic approach within the less rigorous and simpler Eddington-Robertson-Schiff PPN version, it was found the bound $|\bar \nu_\odot| \lesssim 10^{-16}$ for RGGR. Here we find a slightly stronger bound for RGGR,\footnote{Indeed, as argued in Ref.~\cite{Rodrigues:2016tfm}, although the used approach was not as rigorous as the one employed here, the bounds derived on \cite{Rodrigues:2016tfm} should be an order of magnitude estimation.} $|\bar \nu_\odot| \lesssim 10^{-17}$ (both of these bounds consider the Solar System as part of the Milky Way, see Sec.~\ref{subsec:EPE}). Moreover, the present work also address  bounds for a more general class of theories, whose relation between $G$ and the Newtonian potential is given by the expansion \eqref{eq:GU}. The bounds for such class are stated in Table \ref{tab:nu1nu2}. These bounds should be seen with care since they, for technical convenience, do not consider the EPE. Implementations of the EPE, for different RG extensions, are presented in Secs.~\ref{subsec:EPE} and \ref{subsec:EPE2}. In Sec.~\ref{subsec:EPE2}, we explore relations between $G$ and the Newtonian potential that are simple considering the RG motivation, and that do not follow the original RGGR proposal \cite{Rodrigues:2009vf}. In particular, we find a single peculiar case in which the EPE is irrelevant to the observational bound (the case $n=1$ in eq.~\ref{eq:Gxi}).

Renormalization group extensions of GR at the large scales, as presented in several works (some of them are cited in the introduction), constitute a theoretical possibility which demands to be analysed. We add that it is in connection with QFT in curved spacetime and quantum gravity from the asymptotic safety approach. Also, it leads to results and a framework that cannot be naturally achieved by other means.  Among the possible phenomenological consequences, some works have developed on the possibility that perhaps such RG modifications of classical GR may be related to dark matter-like effects \citep[e.g.,][]{Goldman:1992qs, Bertolami:1993mh, Shapiro:2004ch, Moffat:2005si, Reuter:2004nx, Rodrigues:2009vf, Becker:2014jua, Moffat:2015xla}. The latter line of research,  has achieved  interesting nontrivial consequences, but there is not yet an approach sufficiently developed and tested to be clearly better than the standard dark matter approach. Apart from such uncertainties, and on whether one should look for RG effects associated to dark matter or to other effects, the constraints from the Solar System commonly depend on less hypothesis than larger scale phenomena and are commonly of higher precision, hence they should always be taken into consideration.

\acknowledgments
We thank Felipe T. Falciano and Sebasti\~ao Mauro for  remarks on the PPN formalism application, and to Ilya Shapiro for discussions on the RG application to gravity. DCR thanks CNPq and FAPES (Brazil) for partial financial support, AOFA and NB thank CAPES (Brazil) for financial support.

\appendix

\section{Covariant scale setting and new PN potentials} \label{app:covariant}

Here we consider the covariant  extension as proposed in Refs.~\cite{Rodrigues:2015hba, Rodrigues:2016tfm}. Considering the latter references, the scale setting \eqref{eq:ss} has a covariant extension given by
\begin{equation}\label{eq:mufuuh}
	\mu  = f(\Psi),
\end{equation}
with
\begin{equation}
	\Psi \equiv h_{\alpha\beta}u^\alpha u^\beta	\,.
\end{equation}
In the above, $u^\alpha$ is the fluid four-velocity defined in \eqref{emt},  $h_{\alpha \beta} \equiv g_{\alpha \beta} - \gamma_{\alpha \beta}$, and  $\gamma_{\alpha \beta}$ is the reference metric, which we use the Minkowski metric. Hence, the $h_{\alpha \beta}$ that appears in eq.~\eqref{eq:mufuuh} is the same that appears in eq.~\eqref{eq:getah}.

We expand $G^{-1}$ as a power series on $\Psi$, similarly to eq.~\eqref{eq:GU},
\begin{equation}
	G^{-1}=G_f^{-1}+ \sum_{n=1}^{\infty}\sigma_n\Psi^n\,.
\end{equation}
In the above,  $G_f$ is the value of $G$ when $\Psi=0$ (it needs not to coincide with $G_e$ from eq.~\ref{eq:GU}). We use $\sigma_n$ in place of $\nu_n$ to avoid confusion, since these quantities are in general different. Rewriting the field equation \eqref{eqd} up to the first post-Newtonian order, it results
\begin{align}\label{eq:feq}
R_{\alpha\beta}=& \ G_f (1-G_f\sigma_1\Psi) \bigg[8\pi \left(T_{\alpha\beta}-\frac{T}{2}g_{\alpha\beta}\right)+  \nonumber\\[1ex]
& + \sigma_1 \nabla_\alpha\nabla_{\beta}\Psi + \sigma_2 \nabla_\alpha\nabla_{\beta}\Psi^2 +\nonumber\\[1ex]
&  +  \frac 12 g_{\alpha\beta}\sigma_1\Box \Psi + \frac 12 g_{\alpha\beta}\sigma_2\Box \Psi^2  \bigg]\,.
\end{align} 
Using eq.~\eqref{u0}, $\Psi$ is expanded as follows,
\begin{equation}\label{eq:w}
\Psi= h_{00} + h_{00}^2 + {h}_{00}v^2 + 2{h}_{0i}v^i + {h}_{ij}v^i v^j +{\cal O}(v^6)\,.
\end{equation}

The relation between $G_f$ and $G_\mscript{N}$ is found from the Poisson equation $\nabla^2h_{00} = - 8 \pi G_\mscript{N} \rho$ at the Newtonian order, which implies
\begin{equation}
	G_f = \frac{G_\mscript{N}}{1 + \sigma_1 G_\mscript{N}} \, . 	
\end{equation}
In the following, we use $G_\mscript{N}=1$. The relation above is similar to eq.~\eqref{eq:GeGN}, but we stress that $\nu_n$ and $\sigma_n$ are associated to different expansions, threfore their values will  in general be different as well.

Before expressing the metric solution up to the 1PN order, first we solve eq.~\eqref{eq:feq} for $h_{00}$ and $h_{ij}$ up to order $v^2$, and $h_{0i}$ to order $v^3$. In this case, it is sufficient to consider  $\Psi\approx h_{00}$. The procedure is the same one of Sec.~\ref{sec:ppn apply}, and it yields,
\begin{align}\label{eq:2ndO}
 &h_{00}=2U+O(v^4)\,,\\[1ex] &h_{ij}=2\left(1-\frac{2\sigma_1}{1+\sigma_1}\right)U\delta_{ij}+O(v^4)\,,\\[1ex]
 &h_{0i}=-\,\frac{1}{1+\sigma_1}\left(\frac{7}{2}\,V_i+\frac{1}{2}\,W_i\right)+O(v^5)\,.
\end{align}

Now we proceed to obtain $h_{00}$ up to $v^4$ order. In this case, the fourth-order terms that appear in eq.~\eqref{eq:w} do contribute. The resulting expression for $h_{00}$ reads,
\begin{align}\label{eq:h00}
h_{00}=& \ 2U-2\left[1-\frac{\sigma_1+2\sigma_2(1+\sigma_1)}{1+\sigma_1}\right]\,U^2 \ +\nonumber \\
& \ +4\Phi_1+ 4\left(1-\frac{3\sigma_1}{1+\sigma_1} \right)\,\Phi_2+2\Phi_3+6\Phi_4 \ +\nonumber\\
& \ +2\sigma_1Uv^2- 7\sigma_1V_iv^i-\sigma_1W_iv^i + O(v^6)\,.
\end{align}

The standard Will-Nordtvedt PPN formalism \cite{Will:1993ns, Will:2014kxa} does not include the three last terms in eq.~\eqref{eq:h00}.  The above is not the only field equation of the covariant formulation, and neither it is complete, since the $\Lambda$ term was not considered (see Appendix \ref{app:lambda}). Nonetheless, it is sufficient to show that new potentials will appear. In conclusion, the PPN analysis of the covariant extension of the scale setting \eqref{eq:ss}, as proposed in \cite{Rodrigues:2015hba, Rodrigues:2016tfm}, demands an extension of the formalism, including the potentials above, which is beyond the purpose of this work. Theories that are not covered by the PPN formalism are not rare in the literature \cite[e.g.,][]{Clifton:2008jq,Bittencourt:2016smd}.

\section{$\Lambda$ and violation of  asymptotic flatness or analyticity} \label{app:lambda}

In this appendix it is shown that the $\Lambda$ term either violates asymptotic flatness or it cannot be expressed as an analytical function, which are necessary conditions for the application of the PPN formalism. We also comment on the possible physical impact of $\Lambda$ in the Solar System.

The $\Lambda$ term includes a $\Lambda_0$ constant, which reduces to the cosmological constant of GR if $\nu=0$, and RG corrections that depend on the RG scale $\mu$ and on powers of $\nu$. The constant $\Lambda_0$ in GR necessarily leads to non-asymptotically flat spacetimes, hence it is not considered in standard PPN Solar System analysis. This is also physically reasonable since, up to first Post-Newtonian order (1PN), considering its value  as inferred from the cosmological observations, it has negligible impact on the Solar System dynamics \cite[e.g.,][]{Sereno:2006re}. Therefore, as a starting point on the $\Lambda$ contribution analysis up to 1PN, we consider 
\begin{equation}\label{eq:L00}
	\Lambda_0=0 \, .	
\end{equation}

 According to Ref.~\cite{Rodrigues:2015hba}, in any region without matter (i.e., $T_{\mu\nu}=0$), writing $\Lambda$ and $G$ as $\Lambda = \Lambda_0 + O(\nu)$ and $G = G_0 + O(\nu)$, then,
\begin{equation}
	\Lambda = \Lambda_0 G_0 G^{-1} + O(\nu^2)\, .	
\end{equation}
Consequently, in vacuum and using $\Lambda_0=0$, one finds $\Lambda=0 + O(\nu^2)$. 

From the above, one concludes that, within the approximation that the Solar System is composed by point particles representing the Sun and the planets, $\Lambda$ should not have a relevant role up to the 1PN order. This is in accordance in particular with the Laplace-Runge-Lenz vector approach of Refs.~\cite{Farina:2011me, Rodrigues:2016tfm}.

On the other hand, the Will-Nordtvedt PPN approach uses a fluid instead of point particles. This change from particles to fluid may lead to different answers depending on the theory \cite{Will:1993ns}, for instance it may change the value of $\beta$ appreciably. 

As commented in Sec.~\ref{sec:review}, the $\Lambda$ expression as a function of $\mu$ should be derived from the field equations and hence it is not universal (say, in vacuum, inside a star or in cosmology $\Lambda$ may have different dependences on $\mu$). Nonetheless, for a fixed system, the Solar System, $\Lambda$ should be a fixed function of $\mu$. Using the scale setting \eqref{eq:ss} and expanding $\Lambda$ similarly to what was done for $G$ in eq.~\eqref{eq:GU}, let
\begin{equation}\label{TaylorLambda}
\Lambda = \Lambda_{0}+ \sum_{n=1}^\infty \Lambda_{n} U^n\,.
\end{equation}  
The hypothesis in the above is that, although $\Lambda$ is not in general an analytical function, perhaps it can be approximated by one in the Solar System and up to 1PN order. We will show that this hypothesis cannot be true in an asymptotically flat spacetime.

With $\Lambda$,  the field equations \eqref{eqd} become
\begin{align}\label{eqd2}
R_{\mu\nu}= G\bigg[8\pi & \left(T_{\mu\nu}-\frac{1}{2}g_{\mu\nu}\,T\right) \ +  \\
& + \ \nabla_\mu \nabla_\nu \left(G^{-1}\right) + \frac{1}{2}g_{\mu\nu}\Box\left(G^{-1}\right)\bigg]+ \Lambda g_{\mu\nu}\,. \nonumber
\end{align}   

To proceed with the PPN analysis, one needs to find the metric solution up to order $v^4$. As a first step,  the equation for the zeroth order on $v$ contribution leads to eq.~\eqref{eq:L00}, as expected. The next step is to compute the Newtonian limit which means evaluate $h_{00}$ up to order $v^2$. Thus, using $G_\mscript{N}=1$, 
\begin{equation}
h_{00}= 2U-\Lambda_{1}\chi\;,
\end{equation}
which extends eq.~\eqref{eq:h00GU}. The potential $\chi$ is defined in eq.~(\ref{superpotential}). 

According the PPN formalism, the weak field expansion is about Minkowski metric, but $\chi$ is a potential that diverges at infinity and there is no gauge freedom to remove it, therefore,
\begin{equation}
	\Lambda_{1}=0\, .	
\end{equation}
With the above result, any contribution from $\Lambda$ to the metric may appear only at the $v^4$ order or higher.

Since the $\Lambda$ contribution to the field equations is simply an additional term that depends on no derivates, its contribution to the metric can be easily obtained following the same steps used to derive eq.~\eqref{eq:h00rggr}, leading to, up to the terms of order $v^4$,
\begin{align} 
h_{00}=& \ 2U -2\left[1+ \frac{\nu_{1}^2-\nu_{2}(1+\nu_{1})}{(1+\nu_{1})^2}\right]\!U^2 \ +\nonumber\\[1ex]
& \ + \frac{4\Phi_1}{1+\nu_{1}} +  \frac{2\Phi_3}{1+\nu_{1}}+ \frac{6\Phi_4}{1+\nu_{1}} \ +\label{eq:h00rggrL}\\[1ex]
& \ + \left[\frac{4\left(1-\nu_{1}+\nu^2_{1}\right)}{(1+\nu_{1})^2}-\frac{4\nu_{1}}{1+\nu_{1}}\right] \Phi_2 + 2\Lambda_{2}\aleph\,, \nonumber
\end{align}
where $\aleph$ is a new post-Newtonian potential defined as
\begin{equation}
\aleph=-\frac{1}{4\pi}\int\frac{U'^2}{|\bf{x}-\bf{x'}|}d^3\bf{x'}\,.
\end{equation}
The other metric components  are the same as in eq.~(\ref{33}). 

For large distances from the system, $U$ should decay linearly with the distance, and therefore $\aleph$ diverges logarithmically, implying that 
\begin{equation}
	\Lambda_2 = 0,	
\end{equation}
to preserve asymptotic flatness. With above, the contribution from $\Lambda$ is completely eliminated up to the 1PN order.

In conclusion, the $\Lambda$ term cannot be considered within the standard form of the Will-Nordtvedt PPN formalism. We have not proved that its contribution is dynamically negligible, and hence by not considering it one may be inserting violations of energy-momentum conservation that are relevant at 1PN order within the fluid description. However, considering the point particle case, in which $\Lambda$ becomes zero everywhere, it is unlikely that its inclusion can change the derived bounds by orders os magnitude. For instance, in case a full inclusion of $\Lambda$ in the dynamics can lead to $\zeta_2 = 0$, the bound on $\nu_2$ in table \ref{tab:nu1nu2} will change, but hardly by an order of magnitude, in particular since the constraint on $\beta$ is rather close to the constraint that comes from $\zeta_2$.

\bibliographystyle{apsrev4-1} 
%\bibliography{bibdavi2016c}{} 
\bibliography{bibPPNRGGR}{} 

\end{document}